\RequirePackage{ifpdf}
\ifpdf % We are running pdfTeX in pdf mode
\documentclass[pdftex]{sigma}
\else
\documentclass{sigma}
\fi

\def\C{\mathbb C}
\def\R{\mathbb R}
\def\Z{\mathbb Z}

\def\T{\mathbb T}
\def\t{\mathfrak t}

\def\g{\mathfrak g}
\def\r{\rangle}
\def\l{\langle}

\def\z{\check z}

\begin{document}

\allowdisplaybreaks

\renewcommand{\PaperNumber}{076}

\FirstPageHeading

\ShortArticleName{Orthogonality of $C$-, $S$-, and $E$-Functions}

\ArticleName{Orthogonality within the Families of $\boldsymbol{C}$-, $\boldsymbol{S}$-, and\\
$\boldsymbol{E}$-Functions of Any Compact Semisimple Lie Group}

\Author{Robert V. MOODY~$^\dag$ and Jiri PATERA~$^\ddag$}
\AuthorNameForHeading{R.V. Moody and J. Patera}

\Address{$^\dag$~Department of Mathematics,
         University of Victoria,
         Victoria, British Columbia, Canada} 

\EmailD{\href{mailto:rmoody@uvic.ca}{rmoody@uvic.ca}} 
\URLaddressD{\href{http://www.math.ualberta.ca/~rvmoody/rvm/}{http://www.math.ualberta.ca/\~{}rvmoody/rvm/}}

\Address{$^\ddag$~Centre de Recherches Math\'ematiques,
         Universit\'e de Montr\'eal,\\
$\phantom{^\ddag}$~C.P.6128-Centre ville,         Montr\'eal, H3C\,3J7, Qu\'ebec, Canada}
\EmailD{\href{mailto:patera@crm.umontreal.ca}{patera@crm.umontreal.ca}}
\URLaddressD{\href{http://www.crm.umontreal.ca/~patera/}{http://www.crm.umontreal.ca/\~{}patera/}}

\ArticleDates{Received October 30, 2006; Published online November 08, 2006}

\Abstract{The paper is about methods of discrete Fourier analysis in the context of Weyl 
group symmetry. Three families of class functions are def\/ined on the maximal torus 
of each compact simply connected semisimple Lie group $G$. Such functions can 
always be restricted without loss of information to a fundamental region $\check F$ 
of the af\/f\/ine Weyl group. The members of each family satisfy basic orthogonality 
relations when integrated over $\check F$ (continuous orthogonality). 
It is demonstrated that the functions also satisfy discrete orthogonality 
relations when summed up over a f\/inite grid in $\check F$ (discrete orthogonality), 
arising as the set of points in $\check F$ representing the 
conjugacy classes of elements of a f\/inite Abelian subgroup of the maximal torus $\T$.
The characters of the centre $Z$ of the Lie group allow one to 
split  functions $f$ on $\check  F$ into a sum $f=f_1+\cdots+f_c$, 
where $c$ is the order of $Z$, and where the component functions $f_k$ 
decompose into the series of $C$-, or $S$-, or $E$-functions from one congruence class only.}


\Classification{33C80; 17B10; 42C15}

\section{Introduction}

\looseness=1
The purpose of this paper is to prove continuous and discrete 
orthogonality relations  for three families of special functions 
that arise in connection with each compact simple simply connected 
Lie group $G$ of rank $n$. These functions are def\/ined on $\R^n$ 
(more precisely on the Lie algebra of a maximal torus of $G$). They 
are periodic and have invariance properties with respect to the 
af\/f\/ine Weyl group $W_{\rm aff}$ of $G$. Following \cite{P} we call 
them $C$-, $S$-, and $E$-functions to underline the fact that they 
are generalizations of the cosine, sine, and the exponential 
functions.

Group transforms, similar to familiar Fourier and cosine transforms, 
motivate this paper.  The $C$-, $S$-, and $E$-functions serve as 
fundamental functions into which functions with suitable invariance 
properties can be expanded. Their orthogonality properties make the 
expansions easy to perform.

It is discrete transforms that primarily interest us here because 
they have  a number of practically useful properties  
\cite{P1,MP1,MP2,MP3,KMPS,MMP2,GP,PZ1,PZ2,PZ3,KaP,KP,KP2,AP1,AP2,APSA,PZH,GPZ,AGP}. 
In particular, continuous extension of the 
discrete transforms smoothly interpolate digital data in any 
dimension and for any lattice symmetry af\/forded by the structure of 
the given Lie group~$G$. Many examples show that relative to the 
amount of available data, these transforms provide much smoother 
interpolation than the conventional Fourier transform.

At f\/irst encounter with such a theory, one might suspect that  there 
are two major limitations to its applicability: f\/irst that it 
involves symmetry that probably does not exist in most of problems, 
and second that it involves data grids which ref\/lect this symmetry 
and that these may be dif\/f\/icult to f\/ind and laborious to create. In 
fact neither of these is the case.

First regarding the symmetry, suppose that some compact domain $D$  
in $\R^n$ is given and that we are to analyze certain function $f$  
def\/ined on $D$. After enclosing $D$ inside the fundamental region 
$\check F$ of the af\/f\/ine Weyl group $W_{\rm aff}$, we may extend $f$ 
immediately to a $W_{\rm aff}$-invariant function~$\tilde f$ on all 
of $\R^n$. Thus $\tilde f$ is completely periodic and also carries 
full $W$-symmetry. Analysis of $\tilde f$ yields information about 
$f$ by restriction to $\check F$.
Second, with regards to grids, the f\/inite subgroups of the torus 
$\T$  of $G$ provide us with inf\/initely many dif\/ferent $W_{\rm 
aff}$-invariant grids. There is an elegant and rapidly computable 
method of labeling the points of such grids that lie in the 
fundamental region $\check F$ \cite{MP1}.
An additional feature of the connection with the Lie groups  is that 
we may use the central elements to ef\/fect basic splitting of 
functions, similar to the familiar decomposition of a function into 
its even and odd parts.

The purpose of this paper is to record for future reference  the 
exact form of orthogonality relations and central splittings of the 
various classes of the functions ($C$-, $S$-, $E$-) in the context 
of discrete subgroups of $\T$ of $G$. 
The special functions considered here are new in their  intended use 
though the $C$- and $S$-functions are well known in Lie theory. More 
precisely, $C$-functions, which have also been called Weyl orbit 
sums or Weyl orbit functions, are constituents of the irreducible 
characters of $G$. The discrete version of the $C$-transform has 
been described in general in our~\cite{MP1}. Its various 
mathematical applications are found  in \cite{MP2,MP3,KMPS,MMP2,GP,PZ1,PZ2,PZ3,KaP,KP,KP2,AP1,AP2}. For 
other applications, see \cite{AP1,AP2,APSA,PZH,GPZ,AGP}. Existence of the 
continuous version of the transform is implied there as well. A 
review of properties of $C$-functions is in~\cite{KP}.

The $S$-functions appear in the Weyl character  formula \cite{W}, 
where each character is expressed as the ratio of two particular 
$S$-functions. In this paper the functions are put to a dif\/ferent 
use. A detailed description of the four special cases of 
$S$-functions depending on two variables (semisimple Lie groups of 
rank two), is found in \cite{PZ3}. A general review of properties of 
$S$-functions  is in \cite{KP2}.
The $E$-functions are not found in the literature except \cite{P}.  
Their def\/inition is rather natural, once the even subgroup of the 
Weyl group is invoked.

The names $C$-, $S$-, and $E$-functions make allusion  to the fact 
that, for rank $n=1$, the functions become cosine, sine, and the 
exponential functions. The underlying compact Lie groups are $SU(2)$ 
for the $C$- and $S$-functions, and either $SU(2)$ or $U(1)$ for the 
$E$-functions, because $W^e=1$ in the $SU(2)$ case.
From a general perspective, our functions are special  cases of 
functions symmetrized by summing group dependent terms over all 
elements of a f\/inite group. In our cases the f\/inite groups are the 
Weyl group $W$ for $C$ and $S$, and its even subgroup $W^e$ for the 
$E$-functions.

In Section~2 we recall pertinent facts from the Lie theory.  In 
Section~3 orthogonality properties of common exponential function 
are invoked, when the function is integrated over $n$-dimensional 
torus, or when it is summed up over a suitable f\/inite subgroup of 
the torus. In the  subsequent Sections~4--6, the families of $C$-, 
$S$-, and $E$-functions are dealt with respectively. The three 
families are def\/ined for any compact semisimple Lie group. The rank 
of the group is equal to the number of independent variables.  In 
each case the def\/inition is followed by the proof of the continuous 
and discrete orthogonality relations. In  Section~7, it is shown how 
the general problem of decomposition of a function into series of 
$C$-, or $S$-functions can be split into $c$ sub-problems, where $c$ 
is the order of the centre of the Lie group. Central splitting is 
illustrated on the examples of the groups $SU(3)$, $Sp(4)$, and 
$SU(2)\times SU(2)$. Concluding remarks are in the last section.

\section{Preliminaries}

The uniformity of the theory of semisimple Lie groups furnishes us 
with many standard concepts for any compact simply connected simple 
Lie group $G$ of rank $n$. We assume that the reader is familiar 
with the basic theory of such groups but we outline some of it here 
in order to establish the notation. For more details see \cite{KMPS} 
and \cite{MP2}.

We denote by $G$ a compact simply connected simple Lie group of rank 
$n$; $\g$ is the Lie algebra of $G$ and $\g_\C:=\C\otimes_\R\g$ is 
its complexif\/ication. Let $\T$ be a maximal torus of $G$,  
$\dim_\R\T=n$. Let $\t'$ be the Lie algebra of $\T$. Then 
$\t:=i\t'=\sqrt{-1}\t'$ is an $n$-dimensional real space. We have 
the exponential map
\[
\exp2\pi i(\cdot) :\quad\t\ \longrightarrow\  \T.
\]
The kernel $\check{Q}$ of $\exp2\pi i(\cdot)$ is a lattice of rank $n$ in $\t$ called the {\it co-root lattice}.

Let $\t^*$ be the dual space of $\t$ and let 
\[
\l\cdot,\cdot\r :\quad \t^*\times\t
  \ \longrightarrow\ \R
\]
be the natural pairing of $\t^*$ and $\t$.

For any f\/inite dimensional (smooth complex) representation
\[
\pi_V :\quad G\ \longrightarrow\ GL(V),
\]
$V$ decomposes into $\T$-eigenspaces
\[
V=\oplus_{\lambda\in\Omega(V)} V^\lambda,
\]
where $\Omega(V)\subset \t^*$ is a subset of the dual space $\t^*$ of $\t$. 
Then on each weight space $V^\lambda$, $\pi_V(\exp{2\pi ix)}$ is simply 
a multiplication by $e^{2\pi i\l\lambda,x\r},$ for all $x\in\t$. 
The set $\Omega(V)$ is the set of weights of $\pi_V$. Since
\[
\check{Q}=\ker\left(\exp{2\pi i(\cdot)}\right)
      \text{ one has } e^{2\pi i\l\lambda,x\r}=1
      \text{ for all }\lambda\in\Omega(V)
      \text{ and for all } x\in\check{Q}.
\]
Thus $\lambda\in\Omega(V)$ implies that $\lambda$ is in the $\Z$-dual lattice $P$ to $\check{Q}$.
 This is a rank $n$ lattice in $\t^*$ called the {\it weight lattice}. Every $\lambda\in P$
 occurs in $\Omega(V)$ for some representation $\pi_V$. 

For the adjoint representation, the eigenspace decomposition is
\begin{equation}
\g_\C=t_\C\oplus\bigoplus_{\alpha\in\Delta}\g_C^\alpha\,,
\end{equation}
i.e. $\{0\}\cup\Delta=\Omega(\text{adjoint rep.})$ and $\Delta $ is called the {\it root system} 
of $G$ relative to $\T$. The $\Z$-span~$Q$ of $\Delta$ is an $n$-dimensional 
lattice in $\t^*$, called the {\it root lattice}. Its $\Z$-dual in $\t$ is the {\it co-weight lattice} $\check{P}$.

Let us summarize relations between the lattices:
\[
\begin{matrix}
   \t^*  &  &\\
   \cup &  &\\
   P    &\overset{\text{$\Z$-dual}}\longleftrightarrow &                  \check{Q}\\
   \cup &  &\cap\\
   Q    &\overset{\text{$\Z$-dual}}\longleftrightarrow & \check{P}\\
        &  &\cap\\
        &  &\t\\
\end{matrix}
\]
We have the index of connection:
\[
\left|P/Q\right|=\left|\check{P}/\check{Q}\right|
\]
which is the order of the centre of $G$.

Let
\[
(\cdot\;,\cdot) :\quad\t^*\times\t^*\ \longrightarrow\ \R
\]
be the transpose of the Killing form of $\g$ restricted to $\t$. Then we can introduce the following four bases:

$\bullet\quad
\{\alpha_1,\dots,\alpha_n\}$,\quad basis of $Q$ formed by a set of  simple roots,

$\bullet\quad
\{\check{\omega}_1,\dots,\check{\omega}_n\}$,\quad basis of $\check{P}$ dual to the basis $\{\alpha_1,\dots,\alpha_n\}$,

$\bullet\quad
\{\check{\alpha}_1,\dots,\check{\alpha}_n\}$,\quad basis of $\check{Q}\subset\t$ of simple co-roots. They are def\/ined by 
\[
\l\phi,\check{\alpha}\r=\frac{2(\phi,\alpha)}{(\alpha,\alpha)} \qquad\text{for all}\quad \phi\in\t^*,
\]

$\bullet\quad
\{\omega_1,\dots,\omega_n\}$,\quad basis of $P$, the 
fundamental weights dual to  $\{\check{\alpha}_1,\dots,\check{\alpha}_n\}$.
\medskip

For each  $M\in\Z_{>0}$, we introduce two f\/inite subgroups of $\T$. For  $x\in\t$, such that
\[
\left(\exp 2\pi ix\right)^M=1
\]
we have
\[
\exp(2\pi iMx)=1\quad\Longleftrightarrow\quad Mx\in\check{Q}
  \quad\Longleftrightarrow\quad x\in\tfrac1M\check{Q}.
\]
Then $\tfrac1M\check{Q}/\check{Q}$ is isomorphic to the group 
of all elements of $\T$ whose order divides $M$. This is our f\/irst point group.

Then we have the group $\tfrac1M\check{P}/\check{Q}$, of all elements with the property
\[
\exp(2\pi ix)^M\vert_{\text{adj.\,rep.}}=1.
\]
This is the second f\/inite group.

Thus
\begin{equation}\label{first}
\tfrac1M\check{P}/\check{Q}\supset\tfrac1M\check{Q}/\check{Q} 
     \qquad\text{and}\qquad \left [\tfrac1M\check{P}/\check{Q}\right]\ :\
\tfrac1M\check{Q}/\check{Q}]=[\check{P}:\check{Q}]
\end{equation}

For each of these two groups we have natural dual, in the sense of duality in Abelian groups. Namely, we have the map
\begin{alignat}{2}
P&\times \tfrac1M\check{Q}
   &\quad &\longrightarrow\quad M\text{-th roots of 1}\\
(\lambda &\ , \ a)
   &\quad &\mapsto\quad e^{2\pi i\l\lambda,a\r}\notag
\end{alignat}
or equivalently,
\begin{alignat}{2}\label{second}
         P/MP &\times \tfrac1M\check{Q}/\check{Q}
                   &\quad &\longrightarrow\quad U(1)\\
    (\lambda&,a) &\quad &\mapsto\quad e^{2\pi i\l\lambda,a\r} , \notag
\end{alignat}
which is a dual pairing. Likewise
\begin{equation}
P/MQ\times \tfrac1M\check P/\check Q
  \quad\longrightarrow\quad U(1),
\end{equation}
is a dual pairing.

Of course the critical object in this theory is the Weyl group $W$ 
which describes the inherent symmetry of the situation. It acts on 
$\T$, $\t$, and $\t^*$.

All the objects \eqref{first}--\eqref{second} are $W$-invariant.  
In particular the f\/inite groups $\tfrac1M\check Q/\check Q$,  
$\tfrac1M\check P/\check Q$,  $P/MQ$,  $P/MP$  play a fundamental 
role in the introduction of discrete methods into computational 
problems involving class functions on $G$ or $\t$.

We let $P^+$ denote the set of dominant weights, and $P^{++}$ the set of strictly dominant weights.

In the sequel $\Gamma$ will denote some $W$-invariant f\/inite 
subgroup of $\T$. Mostly we have in mind the groups 
$\frac1M\check{Q}/\check{Q}$ and $\tfrac1M\check{P}/\check{Q}$, but 
this is not assumed unless explicitly stated so.
Usually we will denote the elements of $\Gamma$ by elements of $\t$ via the exponential map.

On $\t$, $W$ can be combined with the translation group def\/ined by 
$\check Q$ to give the af\/f\/ine Weyl group $W_{\rm aff}=W\ltimes\check Q$.

A convenient fundamental region $\check F$ for the af\/f\/ine Weyl group 
of a simple compact Lie group~$G$ of rank $n$ is the simplex which 
is specif\/ied by the $n+1$ vertices, 
\[ \check F\quad\text{is the 
convex hull of}\quad \big\{0,\tfrac{\check{\omega}_1}{q_1},
      \tfrac{\check{\omega}_2}{q_2},\dots,
      \tfrac{\check{\omega}_n}{q_n}\big\}\subset\t,
\]
where $q_1,\dots,q_n$ are the coef\/f\/icients of the highest 
root of $G$ relative to the basis of simple roots.  
By def\/inition $\check F$ is closed. In case $G$ is semisimple 
but not simple, its fundamental region is the cartesian product
of fundamental regions of its simple components.

In the case of  $W^e_{\rm aff}$, the af\/f\/ine $W^e$,  
its fundamental region is the union of (any) two adjacent copies of $\check F$ of 
the corresponding af\/f\/ine Weyl group $W_{\rm aff}$.

\section{Basic orthogonality relations}

Let us f\/ix an $n$-dimensional torus $\T$ of a compact simple Lie
group $G$. For $\lambda,\mu\in P$ one has the well known basic
continuous orthogonality relation of the elementary functions:
\begin{equation}\label{basiccont}
\int_\T e^{2\pi i\l\lambda,t\r}\overline{e^{2\pi i\l\mu,t\r}}dt
  =\int_\T e^{2\pi i\l\lambda-\mu,t\r}dt=\delta_{\lambda,\mu}\,,
\end{equation}
assuming $\int_\T 1dt=1$.
\medskip

One has the basic discrete orthogonality relation on the f\/inite
Abelian subgroup $\Gamma$ of $\T$ for $\lambda,\mu\in P$,
\begin{equation}\label{basicdiscrete}
\sum_{a\in\Gamma} e^{2\pi i\l\lambda,a\r}
    \overline{e^{2\pi i\l\mu,a\r}}=
    \begin{cases} |\Gamma|&\text{if \ }
                     \lambda|_{\Gamma}=\mu|_{\Gamma},\\
                 0  &\text{otherwise} .
   \end{cases}
\end{equation}
\medskip

For instance consider \eqref{basicdiscrete}. For any $\lambda\in P$ (in particular 
for $\lambda-\mu$ in \eqref{basicdiscrete}),
\[
\sum_{a\in\Gamma}e^{2\pi i\l\lambda,a\r}\in \C
\]
and for any $b\in\Gamma$,
\[
e^{2\pi i\l\lambda,b\r}\sum_{a\in \Gamma}e^{2\pi i\l\lambda,a\r}
  =\sum_{a\in \Gamma}e^{2\pi i\l\lambda,a+b\r}
  =\sum_{a\in \Gamma}e^{2\pi i\l\lambda,a\r}.
\]
So either $\sum\limits_{a\in \Gamma}e^{2\pi i\l\lambda,a\r}=0$ or $e^{2\pi
i\l\lambda,b\r}=1$. If the former is false, $e^{2\pi
i\l\lambda,b\r}=1$ for all $b\in \Gamma$. Therefore
\[
\sum_{b\in \Gamma} e^{2\pi i\l\lambda,b\r}=|\Gamma|\,.
\]

The proof of \eqref{basiccont} is even simpler.

Let $\lambda,\mu\in P$. We say that they satisfy the {\it separation
condition} on $\Gamma$, if one has
\begin{equation}\label{separation}
w\lambda|_{\Gamma}=w'\mu|_{\Gamma}
   \quad\Longleftrightarrow\quad  w\lambda=w'\mu,
\qquad\text{for any}\quad w,w'\in W.
\end{equation}
(Since $\Gamma$ is $W$-invariant, $w\lambda|_{\Gamma}$ is
unambiguously def\/ined.)

\begin{remark} 
The separation condition \eqref{separation}  avoids the
problem that is usually called aliasing in the discrete
transform literature. A given f\/inite group $\Gamma$ can only
separate a f\/inite number of weights, just as in the discrete
Fourier analysis where a f\/inite number of division points can
only be used with band width limited functions.

In our case $\Gamma$ is typically $\tfrac1M\check{Q}/\check{Q}$
or  $\tfrac1M\check{P}/\check{Q}$ and $\Gamma$ can then separate
all the weights of $P\mod MP$ or all the weights of $P\mod MQ$ respectively.
\end{remark}

\section[$C$-functions]{$\boldsymbol{C}$-functions}

Since $\T=\t/ \check Q$ is a compact Abelian group, its Fourier 
analysis is expressed in terms of objects coming from the lattice 
dual to $\check Q$, namely $P$. Thus Fourier expansions of $\T$ are of the form
\[
\check x \ \mapsto \ \sum_{\lambda\in P}
    a_\lambda e^{2\pi i\l\lambda,\check x\r}.
\]
where $\check x\in \t$. These are literally functions on $\t$ but are ef\/fectively functions on $\T$ since
\[
e^{2\pi i\l\lambda,\check x+\check Q\r}
  =e^{2\pi i\l\lambda,\check x\r}
       \qquad\text{for all}\quad \lambda\in P.
\]

\subsection[Definition of $C$-functions]{Def\/inition of $\boldsymbol{C}$-functions} 

Orbit sums, equivalently $C$-functions, $C_\lambda(t)$ are the
f\/inite sums
\begin{equation}\label{defC}
C_\lambda(x):=\sum_{\lambda'\in W\lambda}
  e^{2\pi i\l\lambda',x\r},
           \qquad \lambda\in\T,\quad x\in\check\T.
\end{equation}
We can assume that $\lambda$ is dominant since 
$C_\lambda(\check x)$ depends on the Weyl group orbit $W\lambda$, not on $\lambda$.

As is well-known, the characters of f\/inite dimensional
representations are linear combinations of $C$-functions
\eqref{defC}. Coef\/f\/icients of such a linear combination are the
multiplicities of the domi\-nant weights in the weight system of
the representation. In general, it is a laborious problem to
calculate the multiplicities. For simple Lie groups of ranks
$n<9$ a limited help is of\/fered by the tabulation \cite {BMP}.

\subsection[Continuous orthogonality of $C$-functions]{Continuous orthogonality of $\boldsymbol{C}$-functions} 

For $\lambda$, $\mu$ dominant, we have the continuous orthogonality
relations:
\begin{equation} 
\int_\T C_\lambda(\check x)\overline{C_\mu(\check x)}d\check x
   =|W|\int_{\check F} C_\lambda(\check x)
    \overline{C_\mu(\check x)} d\check x =
  \begin{cases} |W\lambda|&\text{if } \lambda=\mu,\\
                       0  &\text{otherwise.}
  \end{cases} 
\end{equation}
Here overline denotes complex conjugation, and $|W|$, $|W\lambda|$
respectively stand for the size of the Weyl group and for the size of the orbit containing $\lambda$.

\subsection[Discrete orthogonality of $C$-functions]{Discrete orthogonality of $\boldsymbol{C}$-functions} 

Let $\Gamma$ be a $W$-invariant f\/inite subgroup of $\T$. Let
\[
\check F\cap \Gamma
              =\{\check u_1,\dots,\check u_N\},
\]
which we view as a set of elements of $\t$.

Let $\lambda,\mu\in P$ and assume that they satisfy the
separation condition \eqref{separation} on $\Gamma$. We wish to
prove the orthogonality relation
\begin{equation}\label{Corthog}
\sum_{j=1}^N|W\check u_j|C_\lambda(\check u_j)   
                 \overline{C_\mu(\check u_j)}  
   =\begin{cases}
         |W\lambda||\Gamma|\quad &\text{if}\quad W\lambda=W\mu,\\
           0 \quad &\text{otherwise}.
   \end{cases}
\end{equation}

To see this, note that 
\begin{equation}
\sum_{\check a\in\Gamma} C_\lambda(\check a)
         \overline{C_\mu(\check a)}
   =\sum_{j=1}^N|W\check u_j|C_\lambda(\check u_j)
        \overline{C_\mu(\check u_j)}  
   =\sum_{\lambda'\in W\lambda}\ \sum_{\mu'\in W\mu} \
    \sum_{\check a\in\Gamma} e^{2\pi i\l\lambda'-   
                           \mu',\check a\r}.
\end{equation}

We have from \eqref{basicdiscrete} and \eqref{separation} 
that for $\lambda'\in W\lambda$ and $\mu'\in W\mu$, 
\[
\sum_{\check a\in\Gamma} e^{2\pi i\l\lambda'-\mu',\check a\r}=
  \begin{cases}
       |\Gamma|  &\text{ if } \lambda'=\mu',\\
       0  &\text{ otherwise},
  \end{cases}
\]
from which  \eqref{Corthog} follows.

\section[$S$-functions]{$\boldsymbol{S}$-functions}

\subsection[Definition of $S$-functions]{Def\/inition of $\boldsymbol{S}$-functions} 

\begin{equation}\label{Sdef}
S_\lambda(\check x):=\sum_{\lambda'\in W\lambda}
  \varepsilon(\lambda') e^{2\pi i\l\lambda',\check x\r} ,
\end{equation}
where 
\[
\varepsilon(\lambda')= (-1)^{l(w)}\qquad
   \text{if}\quad \lambda'=w\lambda.
\]
Here $l(w)$ is the length of $w$ as word in the elementary
ref\/lections of $W$.

\begin{remark}
Note that $S_\lambda(\check x)$ is well-def\/ined only if 
$\operatorname{Stab}_W(\lambda)=\{1\}$, otherwise it is ambiguous.  
Thus each sum \eqref{Sdef} has a unique term $\exp{2\pi i\l\lambda',\check x\r}$ 
with $\lambda'$ strictly dominant ($\lambda'\in P^{++}$) and hence the $S_\lambda$ are parametrized by $P^{++}$.
\end{remark}

The $W$-skew-invariance of  $S_\lambda$ also implies that  
$S_\lambda$ vanishes on the boundaries of af\/f\/ine chambers of $\t$.

\subsection[Continuous orthogonality of $S$-functions]{Continuous orthogonality of $\boldsymbol{S}$-functions} 

Let $\lambda$, $\mu$ be strictly dominant integral weights. Then
\begin{align}
\int_\T S_\lambda(\check x)\overline{S_\mu(\check x)}d\check x
  &=\sum_{\lambda'\in W\lambda}\varepsilon(\lambda')
   \sum_{\mu'\in W\mu} \varepsilon(\mu')
   \int_\T e^{2\pi i\l\lambda'-\mu',\check x\r}d\check x\notag\\
  &=\sum_{\lambda'\in W\lambda}\varepsilon(\lambda')
   \sum_{\mu'\in W\mu}\varepsilon(\mu')\,\delta_{\lambda',\mu'}
  =\begin{cases} 
       |W|\  &\text{if}\ \lambda=\mu,\\
        0           &\text{otherwise}.
   \end{cases}
\end{align}

\subsection[Discrete orthogonality of $S$-functions]{Discrete orthogonality of $\boldsymbol{S}$-functions}

Let $\Gamma$ be a f\/inite $W$-invariant subgroup of $\T$, and
let $\lambda,\mu\in P^{++}$. Assume the
separation condition \eqref{separation}.

We prove that   
\begin{equation}\label{Sorthog}
\sum_{\check u_j\in\Gamma\cap\,\operatorname{int}(\check F)}
S_\lambda(\check u_j)\overline{S_\mu(\check u_j)}
  =\begin{cases}
      |\Gamma|\quad &\text{if}\quad \lambda=\mu,\\
       0       &\text{otherwise} .
    \end{cases}
\end{equation}

To see this we note f\/irst that if $\check u_j$ is on the boundary of $\check F$, 
then there is an ref\/lection  $r$ such that 
$r\check u_j=\check u_j$ and
\[
S_\lambda(\check u_j)=S_\lambda(r\check u_j)
         =-S_\lambda(\check u_j)=0,
\]
so there is no contribution from this $\check u_j$.

Now,
\begin{align}
\sum_{j=1}^n |W\check u_j|S_\lambda(\check u_j)
            \overline{S_\mu(\check u_j)}
   &=\sum_{\check u_j\in\Gamma\cap\,\operatorname{int}(\check F)}
             |W\check u_j|S_\lambda(\check u_j)
             \overline{S_\mu(\check u_j)}\notag\\
   &=|W|\sum_{\check u_j\in\Gamma\cap\,\operatorname{int}(\check F)}
                         S_\lambda(\check u_j)
           \overline{S_\mu(\check u_j)}.\label{orto3}
\end{align}

At the same time,
\begin{equation}
\sum_{j=1}^n |W\check u_j|S_\lambda(\check u_j)
                      \overline{S_\mu(\check u_j)}
   =\sum_{\check a\in \Gamma} S_\lambda(\check a)
                     \overline{S_\mu(\check a)}
\end{equation}
and
\begin{align}
\sum_{a\in \Gamma} S_\lambda(\check a)
                             \overline{S_\mu(\check a)}
  &=\sum_{\check a\in \Gamma}\ \  \sum_{\lambda'\in W\lambda} 
   \varepsilon(\lambda')e^{2\pi i\l\lambda',\check a\r}
   \sum_{\mu'\in W\mu}\ \ 
   \varepsilon(\mu')e^{-2\pi i\l\mu',\check a\r}\notag\\
  &=\sum_{\lambda'\in W\lambda}\ \ \sum_{\mu'\in W\mu} 
       \varepsilon(\lambda')\varepsilon(\mu')
    \sum_{\check a\in \Gamma}e^{2\pi i\l\lambda'-\mu',\check a\r}\notag\\
  &=\begin{cases}
       |W||\Gamma|\quad &\text{if}\quad \lambda=\mu,\\
        0          &\text{if}\quad \lambda\neq\mu .
    \end{cases}\label{orto4}
\end{align}

Putting together \eqref{orto3} and \eqref{orto4}, we obtain 
\eqref{Sorthog}.

\section[$E$-functions]{$\boldsymbol{E}$-functions}

\subsection[Definition of $E$-functions]{Def\/inition of $\boldsymbol{E}$-functions} 

Let $W^e\subset W$ be the even subgroup of $W$, and
$W^e_{\operatorname{aff}}$ the even subgroup of
$W_{\operatorname{aff}}$,
\[
W^e:=\{w\in W\mid (-1)^{l(w)}=1\},\qquad
W^e_{\rm aff}
    :=W^e\ltimes\check Q.
\]

Let $r$ be any f\/ixed simple af\/f\/ine ref\/lection, and set
\[
\check F^e:= \check F\cup r\check F.
\]
Then $F^e$ is a fundamental region for
$W^e_{\operatorname{aff}}$, the af\/f\/ine even Weyl group. Let
\[
P^+_e:= P^+\cup rP^+.
\]

For $\lambda\in P$,
\begin{equation}
E_\lambda(\check x):=\sum_{\lambda'\in W^e\lambda}
  e^{2\pi i\l\lambda',\check x\r} .
\end{equation}

Since $E_\lambda(\check x)$ depends on $W^e\lambda$, not on $\lambda$,
we can suppose $\lambda\in P^+_e$.
\begin{equation}
C_\lambda=
 \begin{cases}
    E_\lambda+E_{r\lambda}\quad&\text{if}\ \lambda\neq r\lambda,\\
    E_\lambda                  &\text{if}\ \lambda=r\lambda,
 \end{cases}
\end{equation}
which also shows that $E_{r\lambda}$ depends only on $\lambda$, not on the choice of the ref\/lection $r$. 

Let $\Gamma$ be as above. Put
\[
\Gamma\cap \check F^e=\{\check u_1^e,\check u_2^e,\dots,\check u_m^e\}.
\]

\subsection[Continuous orthogonality of $E$-functions]{Continuous orthogonality of $\boldsymbol{E}$-functions} 

Let $\lambda,\mu\in P^+_e$,
\begin{align}
\left(E_\lambda,E_\mu\right)
  &=\int_\T E_\lambda(\check x)\overline{E_\mu(\check x)}d\check x
   =\sum_{\lambda'\in W^e\lambda}\ \ \sum_{\mu'\in W^e\mu}
    \int_\T e^{2\pi i\l\lambda'-\mu',\check x\r}d\check x \notag\\
  &=\sum_{\lambda'\in W^e\lambda}\ \ \sum_{\mu'\in W^e\mu}
    \delta_{\lambda',\mu'}
  =\begin{cases}
     |W^e\lambda|\quad&\text{if}\quad\lambda=\mu,\\
      0                &\text{otherwise}.
   \end{cases} 
\end{align}

\subsection[Discrete orthogonality of $E$-functions]{Discrete orthogonality of $\boldsymbol{E}$-functions} 

Assume $\lambda,\mu\in P^+_e$, and assume the separation
condition \eqref{separation} for the group $W^e$.
Since $E_\lambda$, $E_\mu$ are $W^e$-invariant,
\begin{align}
\sum_{j=1}^m |W^e\check u_j|E_\lambda(\check u_j^e)
              \overline{E_\mu(\check u_j^e)}
  & =\sum_{a\in \Gamma} E_\lambda(\check a)\overline{E_\mu(\check a)}\notag\\
 &=\sum_{\lambda'\in W^e\lambda}\ 
    \ \sum_{\mu'\in W^e\mu}
    \ \sum_{\check a\in \Gamma}e^{2\pi i\l\lambda'-\mu',\check a\r}
    =|\Gamma|\sum_{\lambda'\in W^e\lambda}\ \ \sum_{\mu'\in W^e\mu}
     \delta_{\lambda',\mu'}\notag \\
 &=\begin{cases}
     |\Gamma||W^e\lambda|\quad&\text{if}\quad\lambda=\mu,\\
      0                &\text{otherwise} .
         \end{cases} 
\end{align}

\section{Central splitting of the functions}

All but three compact simply connected simple Lie groups, $G_2$, $F_4$, $E_8$, 
have the centre $Z$ of order $|Z|=c>1$.  Consider such a group with $c>1$. Then $Z\subset\T$ and
\[
Z=\{\z_1,\dots,\z_c\},\quad
   \text{where}\quad
c=\left|\check P/\check Q\right|\quad\text{(index of
                                               connection),}
\]
and again we identify the elements of $Z$ with elements of $\t$.

Let $\chi_1,\dots,\chi_c,$ be the irreducible characters of $Z$,
i.e. the homomorphisms
\[
\chi :\quad Z\ \longrightarrow\ U(1).
\]

Each $\lambda\in P$ determines an irreducible character on $Z$,
\begin{gather}
\chi_\lambda \ :\quad\z\ \mapsto\  e^{2\pi
                   i\l\lambda,\z\r}
\end{gather}                   
so,
\begin{gather}
\chi_\lambda  =\chi_j\quad\text{for some}\quad
 1\leq j\leq c,
\end{gather}
and $j$ is called the congruence class of $\lambda$. It depends only on $\lambda\mod Q$. 
Hence it is constant on the $W$- and
$W^e$-orbits of $\lambda$.

The following arguments can be made for $C$-, $S$-, and
$E$-functions.  Using just the $C$-functions, we have
\begin{equation}\label{zact}
C_\lambda(\check x+\z)
  =\sum_{\lambda'\in W\lambda}e^{2\pi i\l\lambda',\check x+\z\r}
  =\chi_j(\z)C_\lambda(\check x)\qquad\text{for all}\quad 
        \z\in Z,
\end{equation}
where $j$ is the congruence class of $\lambda$. Thus for any class
$k$ we have
\[
\sum_{\z\in Z}\overline{\chi_k(\z)}C_\lambda(x+\z)=
  \left(\sum_{\z\in Z}
  \overline{\chi_k(\z)}\chi_j(\z)\right)C_\lambda(x)
  =c\,\delta_{kj}C_\lambda(x).
\]

From this, if $f$ is a linear combination of $C$- or
$S$-functions, we can determine its splitting into the sum of
congruence class components (central splitting) as follows,
\begin{gather}\label{split}
f(x)=f_1(x)+\cdots+f_c(x),
\end{gather}
where
\begin{gather}
f_k(x)=\tfrac1c\sum_{\z\in Z}\overline{\chi_k(\z)}f(x+\z),
          \qquad 1\leq k\leq c.
\end{gather}

\subsection[Example $SU(3)$]{Example $\boldsymbol{SU(3)}$} 

The centre $Z$ has 3 elements, $\{\z_0,\z_1,\z_2\}$ (cyclic group of 3 elements). Its characters are
\begin{align}
\chi_0 &:\quad \z\mapsto 1\quad\text{(trivial character)},
                                         \notag\\
\chi_1 &:\quad \z_1\mapsto  e^{2\pi i/3},\\
\chi_2 &:\quad \z_2\mapsto  e^{-2\pi i/3}.  \notag
\end{align}

According to \eqref{split} a function $f$ on $\check F$ is decomposed into the sum of three functions,
\begin{equation}\label{A2decompose}
f=f_0+f_1+f_2,
\end{equation}
where
\begin{gather}
f_0(\check x)=\tfrac13\{f(\check x)+f(\check x+\z_1)+f(\check x +\z_2)\}, \notag\\
f_1(\check x)=\tfrac13\{f(\check x)+e^{-2\pi i/3}f(\check x+\z_1)+ e^{2\pi i/3}f(\check x+\z_2)\},\label{A2components}\\ 
f_2(\check x)=\tfrac13\{f(\check x)+e^{2\pi i/3}f(\check x +\z_1)+e^{-2\pi  i/3}f(\check x+\z_2)\}\,.\notag
\end{gather}
By substitution of \eqref{A2decompose} into \eqref{A2components}, 
followed by use of \eqref{zact}, one verif\/ies directly the basic 
property of the central splitting  \eqref{A2decompose} of $f(\check x)$: 

\smallskip
{\it Each of the components $f_0,f_1,f_2$ of $f$ decomposes 
into a linear combination of $C$- or $S$-functions from one congruence 
class only, respectively class $0$, $1$, and $2$.}

\smallskip

For specif\/ic application of \eqref{A2components} using a
function $f(\check x)$ given on $\check F$, it is useful 
to make one more step, namely to express each component 
function as a sum of three terms, each term being the given function evaluated at
particular points in $\check F$. For $\check x\in \check F$, the points
$\check x+\z_1$ and $\check x+\z_2$ need not be in $\check F$ 
and need to be brought there by suitable transformations from 
the af\/f\/ine Weyl group of $A_2$.

The vertices of $\check F=\{0,\check\omega_1,\check\omega_2\}$ 
also happen to be the points representing the elements of the 
centre of $SU(3)$, i.e.\ $\z_1=\check\omega_1$, $\z_2=\check\omega_2$, 
and $\z_0=0$ standing for the identity element of $SU(3)$. 
Required transformations are accomplished, for example by the 
following pairs of af\/f\/ine ref\/lections. Putting 
$\check x=a\check\omega_1+b\check\omega_2=(a,b)$, we have
$\check x+\z_1=(a+1,b)$ and $\check x+\z_2=(a,b+1)$,
\begin{gather}
R_{\alpha_1+\alpha_2}R_{\alpha_1}(a+1,b) =(1-a-b,a)\in \check F,\\
R_{\alpha_1+\alpha_2}R_{\alpha_2}(a,b+1) =(b,1-a-b)\in \check F.
\end{gather}
Here $R_\beta$ denotes the af\/f\/ine ref\/lection in the plane 
orthogonal to $\beta$ and passing through the point $\tfrac12\beta$.

Finally one has,
\begin{gather}
f_0(a,b)=\tfrac13\{f(a,b)+f(1-a-b,a)+f(b,1-a-b)\},\notag\\
f_1(a,b)=\tfrac13\{f(a,b)+e^{-2\pi i/3}f(1-a-b,a)+e^{2\pi
          i/3}f(b,1-a-b)\},\label{A2comp}\\ 
f_2a,b)=\tfrac13\{f(a,b)+e^{2\pi i/3}f(1-a-b,a)+e^{-2\pi
          i/3}f(b,1-a-b)\}.\notag
\end{gather}

\subsection[Example $Sp(4)$]{Example $\boldsymbol{Sp(4)}$}

The fundamental region is the isosceles triangle $\check F$ with vertices 
\[
\{0,\tfrac12\check\omega_1,\check\omega_2\}
\]
the right angle being at $\tfrac12\check\omega_1$.
The centre $Z$ has 2 elements. Their parameters inside $\check F$ are
the points $\{0,\check\omega_2\}$.  According to
\eqref{split}, a function $f(\check x)$ on the triangle $\check F$ is
decomposed into the sum of two functions,
$f(\check x)=f_0(\check x)+f_1(\check x)$, where
\begin{gather}
f_0(\check x)=\tfrac12\{f(\check x)
             +f(\check x+\check\omega_2)\},\notag\\
f_1(\check x)=\tfrac12\{f(\check x)
             -f(\check x+\check\omega_2)\}.\label{C2components}
\end{gather} 
When $ \check x\ne 0$, we have $\check x+\check\omega_2$ 
outside of $\check F$. An appropriate af\/f\/ine ref\/lection brings 
$\check x+\check\omega_2$ into~$\check F$. 

A point $\check x
   =x_1\check\omega_1+x_2\check\omega_2=(x_1,x_2)$ 
is in $\check F$, if $0\leq 2x_1+x_2\leq1$. Then 
$\check x+\check\omega_2$ is ref\/lected into~$\check F$ as follows:
\[
R_{2\check\omega_2}(x_1,1+x_2)=(x_1,1-2x_1-x_2).
\]
Therefore the component functions can be written with their
arguments given explicitly in the $\check\omega$-basis:
\begin{gather}
f_0(x_1,x_2)=\tfrac12\{f(x_1,x_2)+f(x_1,1-2x_1-x_2)\},\notag\\
f_1(x_1,x_2)=\tfrac12\{f(x_1,x_2)-f(x_1,1-2x_1-x_2)\}.\label{C2comp}
\end{gather}
The functions \eqref{C2comp} have def\/inite symmetry
properties with respect to the line connecting the points
$\tfrac12\check\omega_1$ and $\tfrac12\check\omega_2$ on the
boundary of $\check F$. In case one has $f(\check x)$ expanded in terms 
of $C$-functions ($S$-functions), $f_0(x_1,x_2)$ 
is symmetric (antisymmetric), while $f_1(x_1,x_2)$ is antisymmetric
(symmetric) with respect to that line.

Equivalently, in case $f(\check x)$ is expanded in terms of
$C$-functions, then $f_0(x_1,x_2)$ is expanded in terms of
$C$-functions of congruence class $0$, while
$f_1(x_1,x_2)$ is expanded in terms of $C$-functions of
congruence class $1$. Similarly, in case the expansion of
$f(\check x)$ is in terms of $S$-functions, 
the previous conclusions about congruence classes pertinent to $f_0$ and $f_1$ expansions are interchanged. 

Recall that the functions $C_\lambda$ and also $S_\lambda$,
where $\lambda=a\omega_1+b\omega_2\in P$, are said to be of
congruence class $0$ or $1$, according to the value of $a\mod2$.

\subsection[Example $SU(2)\times SU(2)$]{Example $\boldsymbol{SU(2)\times SU(2)}$} 

The centre of  $SU(2)\times SU(2)$ is the product $Z=Z_2\times Z_2$ of two cyclic groups 
of order 2. Its order is $c=4$. In $\omega$-basis the parameters of 
the centre elements are the vertices of the square $\check F$:
\[
z_1=(0,0),\quad
z_2=(1,0),\quad
z_3=(0,1),\quad
z_4=(1,1).
\]

Suppose a function $f(x,y)$ (`the data') 
is given within the square $\check F$, and that our goal 
is to consider expansions of $f(x,y)$ into series of $C$- and $S$-functions of $SU(2)\times SU(2)$.

A function $f(x,y)$ on $\check F$ has $0\leq x,y\leq1$. We decompose
$f(x,y)$  into the sum of four component functions which we label by two integers $\mod2$,
\begin{equation}\label{A1A1decompose}
f(x,y)=f_{00}(x,y)+f_{10}(x,y)+f_{01}(x,y)+f_{11}(x,y),
\end{equation}
where
\begin{gather}
f_{00}(x,y)=\tfrac14\{f(x,y)+f(x+1,y)+f(x,y+1)+f(x+1,y+1)\},\notag\\
f_{10}(x,y)=\tfrac14\{f(x,y)-f(x+1,y)+f(x,y+1)-f(x+1,y+1)\},\notag\\
f_{01}(x,y)=\tfrac14\{f(x,y)+f(x+1,y)-f(x,y+1)-f(x+1,y+1)\},\label{decomp}\\
f_{11}(x,y)=\tfrac14\{f(x,y)-f(x+1,y)-f(x,y+1)+f(x+1,y+1)\}.\notag
\end{gather}
The coef\/f\/icients in \eqref{decomp} are taken from the rows of the
$Z_2\times Z_2$ character table:
\[
\begin{array}{rrrr}1&1&1&1\\1&-1&1&-1\\1&1&-1&-1\\1&-1&-1&1
\end{array}
\]

Since the displaced points in \eqref{decomp} are outside of $\check F$, we
bring them back by the appropriate af\/f\/ine ref\/lection. Finally we get
\begin{gather}
f_{00}(x,y)=\tfrac14\{f(x,y)+f(1-x,y)+f(x,1-y)+f(1-x,1-y)\},\notag\\
f_{10}(x,y)=\tfrac14\{f(x,y)-f(1-x,y)+f(x,1-y)-f(1-x,1-y)\},\notag\\
f_{01}(x,y)=\tfrac14\{f(x,y)+f(1-x,y)-f(x,1-y)-f(1-x,1-y)\},\label{A1A1components}\\
f_{11}(x,y)=\tfrac14\{f(x,y)-f(1-x,y)-f(x,1-y)+f(1-x,1-y)\}.\notag
\end{gather}
In \eqref{A1A1components} each component function $f_j(x,y)$ is given
by the values of $f(x,y)$ at four points in $\check F$. Moreover, its $C$-
or $S$-transforms involve the $C$- or $S$-functions of one of the four
congruence classes only. 

\section{Concluding remarks}

\noindent{\bf 1.}\ 
Specif\/ic examples of orthogonalities of $C$-, $S$-, or $E$-functions 
can be found elsewhere in the literature, for example in \cite{AP2,PZ1,PZ2,PZ3,KaP,KP,KP2}.

\smallskip
\noindent{\bf 2.}\ A practically important distinction between the the $C$-, $S$-, and
$E$-transforms comes from the behaviour of the functions of each
family at the $(n-1)$-dimensional boundary $\partial\check F$ of $\check F$. 
The behaviour is shared by all members of the family and it does 
not depend on the type of the underlying Lie group. It is easy to 
 from  the def\/initions that $C$-functions have normal derivative at
  $\partial\check  F$ equal to zero; the $S$-functions 
  are antisymmetric on both sides of  $\partial \check F$, 
  passing trough 0 at  $\partial \check F$; while no such  property is implied in the case of $E$-functions.

\smallskip

\noindent{\bf 3.}\ Besides the $C$-, $S$-, and $E$-transforms, we have been
considering so far, there are other curious possibilities of their
combinations when $G$ is semisimple but not simple. On some
occasions such hybrid transforms may prove to be useful.

 Suppose $G=G_1\times G_2$, where $G_1$ and $G_2$ are simple,
their fundamental regions being $\check F_1$ and~$\check F_2$ respectively. 
A function $f(\check x_1,\check x_2)$ on $\check F_1\times \check F_2$, 
with $\check x_1\in \check F_1$ and  $\check x_2\in  \check F_2$, can be expanded 
using, say, $C$-function on $\check F_1$ and
any of the three types on $\check F_2$. Thus one may have $CS$-, or $CE$-,
 or $SE$-transforms in such a case, rather than $CC$-, $SS$-, 
 or $EE$-transforms we implied to have in the main body of the paper.

\smallskip

\noindent{\bf 4.}\ It is interesting and useful to know when the $C$-, $S$-, 
$E$-functions are real valued. Traditional Lie theory has a ready answer to such a question:

When the opposite involution $w_{\rm opp}$ (the longest element in 
$W$ in terms of the generating ref\/lections $r_{\alpha_1},\dots,r_{\alpha_m}$) 
is the $-1$ map, every $W$-orbit contains with every weight $\mu$ also $-\mu$. 
Then the $C$-functions are real. When $w_{\rm opp}=-1$ and it is even, 
the same goes for $W^e$: every $W^e$-orbit contains with each 
weight $\mu$ also $-\mu$. Then $S$- and $E$-functions are real.

More interesting, perhaps, is when $S$-functions are purely imaginary. 
This happens when $w_{\rm opp}=-1$ and $w_{\rm opp}$ is odd.

 $w_{\rm opp}=-1$ and $w_{\rm opp}$ is  even for
$B_{2k}$, $C_{2k}$, $D_{2k}$, $E_8$, $F_4$, $G_2$. Hence they
have  real valued $C$-, $S$-, $E$-functions.
       
 $w_{\rm opp}=-1$ and $w_{\rm opp}$ is  odd for
       $A_1$, $B_{2k+1}$, $C_{2k+1}$, $E_7$. 
They have real valued $C$-functions, and purely imaginary valued        $S$-functions.

\smallskip

\noindent{\bf 5.}\ A simple straightforward motivation for central splitting of 
a function $f(\check x)$ on $\check F$ of $G$ is in reducing a larger 
problem, decomposition of $f(\check x)$ into its $C$-, or $S$-, 
or $E$-series, into $c$ smaller problems for each component function. 
That alone should lead to some computational economy. More ambitiously, 
one can view the central splitting, described in Section 7, 
as the f\/irst step of a~general multidimensional fast transform. 
This will be described in \cite{MP5}. In the case $G=SU(2)$, 
this transform coincides with the fast cosine and sine transforms.

\smallskip

\noindent{\bf 6.}\ 
For a given group $G$, the $C$- and $S$-functions have 
a number of other interesting properties, most of which are found in~\cite{KP,KP2}. 
 In addition to that, let us point out also the arithmetic properties 
 described in \cite{MP2} for the $C$-functions but apparently never 
 studied for $S$-functions. As an illustration, let us mention that for 
 any $G$ there are only f\/initely many points in $\check F$, 
 at which  the $C$-functions $C_\lambda(\check x)$ of $G$ take integer values for all $\lambda\in P$.

\subsection*{Acknowledgements}

Work supported in part by the Natural Sciences and Engineering Research Council of Canada, MITACS, 
M.I.N.D. Institute of Costa Mesa, Calif., and by Lockheed Martin Canada.

\LastPageEnding

\newpage

\begin{thebibliography}{14}
\footnotesize

\bibitem{P}
 Patera J., Compact simple Lie groups and their $C$-, $S$-, 
and $E$-transforms, {\it  SIGMA}, 2005, V.1, Paper 025, 6 pages, 
\href{http://arxiv.org/abs/math-ph/0512029}{math-ph/0512029}.
 
\bibitem{P1}
 Patera J., Orbit functions of compact semisimple Lie groups as special functions, in Proceedings of Fifth International
Conference ``Symmetry in Nonlinear Mathematical Physics'' (June
23--29, 2003, Kyiv), Editors A.G.~Nikitin, V.M.~Boyko,
R.O.~Popovych and I.A.~Yehorchenko, {\it Proceedings of Institute
of Mathematics}, Kyiv, 2004, V.50, Part~3, 1152--1160.
 

\bibitem{MP1}
 Moody R.V.,  Patera J., Computation of character decompositions of class  functions 
 on compact semisimple Lie groups, {\it  Math. Comp.}, 1987, V.48, 799--827.

\bibitem{MP2}
Moody R.V.,  Patera J., Characters of elements of f\/inite order in simple Lie groups, 
{\it SIAM J. Algebraic Discrete Methods}, 1984, V.5, 359--383.

\bibitem{MP3}
Moody R.V.,  Patera J.,  Elements of f\/inite order in Lie groups and their applications,
in Proceedings XIII International Colloquium on Group Theoretical Methods in Physics 
(College Park, 1984), Editor W.~Zachary, 
 Singapore, World Scientif\/ic Publishers, 1984, 308--318.

\bibitem{KMPS} 
 Kass S., Moody R.V.,  Patera J., Slansky R., Af\/f\/ine Lie algebras,
 weight multiplicities, and branching rules,
  Vol.~I and~II,  {\it Los Alamos Series in Basic and Applied Sciences},  Berkeley,
  University of California Press, 1990.

\bibitem{MMP2} 
McKay W.G., Moody R.V., Patera J., Tables of $E_8$  characters and 
decomposition of plethysms, in Lie Algebras and  Related Topics,  Editors
D.J.~Britten,  F.W.~Lemire and R.V.~Moody, Providence, RI, Amer. Math. Society, 1985, 227--264.

\bibitem{GP} 
 Grimm S., Patera J., Decomposition of tensor products of the fundamental  
representations of $E_8$, in Advances in Mathematical Sciences --  CRM's 25 Years, Editor L.~Vinet, {\it CRM
Proc. Lecture Notes}, Vol.~11,  Providence, RI, Amer. Math. Soc.,  1997, 329--355.

\bibitem{PZ1}
 Patera J., Zaratsyan A., Discrete and continuous cosine 
transform generalized to the Lie groups $SU(3)$ and $G(2)$,  {\it J.~Math. Phys.}, 2005, V.46, 113506, 17 pages.

\bibitem{PZ2}
Patera J., Zaratsyan A., Discrete and continuous cosine transform 
generalized to the Lie groups $SU(2)\times SU(2)$ and $O(5)$,  {\it J.~Math. Phys.}, 2005, V.46, 053514, 25 pages.

\bibitem{PZ3}
Patera J., Zaratsyan A., Discrete and continuous sine transform 
generalized to semisimple Lie groups of rank two,  {\it J.~Math. Phys.}, 2006, V.47, 043512, 22 pages.

\bibitem{KaP}
 Kashuba I., Patera J., Discrete and continuous $E$-transforms 
of semisimple Lie group of rank two, to appear.

\bibitem{KP}
 Klimyk A., Patera J., Orbit functions, {\it SIGMA}, 2006, V.2, Paper 006, 60 pages,  
 \href{http://arxiv.org/abs/math-ph/0601037}{math-ph/0601037}.

\bibitem{KP2}
Klimyk A., Patera J., Antisymmetric orbit functions, {\it SIGMA}, 2007, V.3, to appear.

\bibitem{AP1}
 Atoyan A., Patera J., Continuous extension of  the discrete cosine transform, and its  
    applications to data processing,  
    in   Group Theory and Numerical Analysis,  {\it CRM Proc. Lecture Notes}, Vol.~39,  Providence, RI, 
 Amer. Math. Soc., 2005, 1--15.

\bibitem{AP2}
Atoyan A., Patera J., Properties of continuous Fourier extension of the discrete cosine 
transform and its multidimensional generalization, {\it J.~Math. Phys.}, 2004, V.45, 2468--2491,
\href{http://arxiv.org/abs/math-ph/0309039}{math-ph/0309039}.

\bibitem{APSA}
 Atoyan A., Patera J., Sahakian V., Akhperjanian A., 
Fourier transform method for imaging atmospheric Cherenkov telescopes, {\it Astroparticle Phys.},
2005, V.23, 79--95, \href{http://arxiv.org/abs/astro-ph/0409388}{astro-ph/0409388}.

\bibitem{PZH} Patera J., Zaratsyan A., Zhu H.-M., New class
of interpolation methods based on discretized Lie group
transforms, in SPIE Electronic Imaging (2006, San Jose), 2006, 6064A-06, S1.

\bibitem{GPZ}
Germain M., Patera J., Zaratsyan A., Multiresolution
analysis of digital images using the continuous extension of
discrete group transforms, in SPIE Electronic Imaging (2006, San
Jose),  2006, 6065-03, S2.

\bibitem{AGP} 
Germain M., Patera J., Allard Y.,  Cosine transform
generalized to Lie groups $SU(2)\times SU(2)$, $O(5)$, and $SU(2)\times SU(2)\times SU(2)$:
 application to digital image processing, {\it Proc. SPIE}, 
 2006, V.6065, 387--395. %, Computational Imaging IV, Feb 2006. 

\bibitem{W}
 Humphreys J., Introduction to Lie algebras and representation theory,  New York, Springer, 1972.

\bibitem{BMP}
Bremner M., Moody R.V., Patera J., Dominant weight multiplicities, New York,  Marcel Dekker, 1990.

\bibitem{MP5}
 Moody R.V., Patera J., Group theory of multidimensional fast Fourier transforms, in preparation.

\end{thebibliography}
\end{document}